\documentclass[preprint]{revtex4}
\usepackage[xdvi,dvips]{graphicx}
\usepackage[dvips]{color}
\usepackage{amssymb}
\newcommand{\Prob}{\mbox{Prob}}
\newcommand{\sgn}{\mbox{sgn}}
\renewcommand{\vec}[1]{\mbox{\boldmath $#1$}}

\begin{document}
%%%
\title{Bifurcation analysis in an associative memory model}

\author{Masaki Kawamura}
\affiliation{Faculty of Science, Yamaguchi University \\
 Yoshida 1677-1, Yamaguchi, 753-8512, Japan}

\author{Ryuji Tokunaga}
\affiliation{Institute of Information Sciences and Electronics,
 University of Tsukuba, Tennodai 1-1-1, Tsukuba, 305-8573, Japan}

\author{Masato Okada}
\affiliation{Laboratory for Mathematical Neuroscience, RIKEN Brain
 Science Institute, Saitama, 351-0198, Japan}
\affiliation{"Intelligent Cooperation and Control", PRESTO, JST,\\
    c/o RIKEN BSI, Saitama, 351-0198, Japan}

\date{\today}
%%%

\begin{abstract}
 We previously reported the chaos induced by the frustration of
 interaction in a non-monotonic sequential associative memory model,
 and showed the chaotic behaviors at absolute zero.
 We have now analyzed bifurcation in a stochastic system, namely a
 finite-temperature model of the non-monotonic sequential associative
 memory model.
 We derived order-parameter equations from the stochastic
 microscopic equations.
 Two-parameter bifurcation diagrams obtained from those equations
 show the coexistence of attractors, which do not appear at absolute
 zero, and the disappearance of chaos due to the temperature effect.
\end{abstract}

%\begin{keyword}
% frustration-induced chaos \sep stochastic system \sep bifurcation 
% \sep associative memory 
% \PACS 02.70.Rr \sep 05.45.-a \sep 05.70.Ln
% %\pacs{02.70.Rr}{General statistical methods}
% %\pacs{05.45.-a}{Nonlinear dynamics and nonlinear dynamical systems}
% %\pacs{05.70.Ln}{Nonequilibrium and irreversible thermodynamics}
%\end{keyword}

\maketitle
%%%

\section{Introduction}

Chaos occurs in systems that consist of chaotic or
binary units. For instance, the globally coupled map
and chaos neural networks
\cite{Kaneko1989,Aihara1990,Adachi1997,Shibata1998} consist of chaotic
units, while neural networks consist of non-chaotic units. 
Although the processing units in neural networks are simple binary
units, chaotic behavior can be observed at the macroscopic level. 
Chaotic behavior can be induced by various mechanisms: 
synaptic pruning, synaptic delay, thermal noise, sparse connections,
and/or so on \cite{FukaiShiino1990,Nara1992,vanVreeswijk,Kawamura2003}.
These models are deterministic systems. Chaotic behavior can also be
observed in stochastic systems \cite{Tsuda1992}.  
Using the Dale hypothesis, Fukai and Shiino \cite{FukaiShiino1990} showed
that chaos can occur in neural networks.

Stochastic behavior can be distinguished from chaotic behavior based on
the exponents, e.g., the Lyapunov exponent. Therefore, analyzing 
deterministic chaos in stochastic systems is very interesting
\cite{MatsumotoTsuda1983,ShibataChawanyaKaneko1999}.  While there is a
close relationship between microscopic behavior and macroscopic behavior,
the macroscopic state cannot always be estimated from the microscopic
state.  Frustration-induced chaos is an example of that.
For continuous systems, chaotic behaviors in
some small networks with frustration of interaction can be analyzed at
the microscopic level \cite{Bersini1995,Bersini1998,Bersini2002}.
In some large random neural networks, a dynamical mean-field theory was
introduced to analyze chaotic behaviors by Sompolinsky et
al. \cite{Sompolinsky1988}. 
We have shown that chaos can be induced by
frustration of interaction in a non-monotonic sequential associative
memory model \cite{Kawamura2003}.

The sequential associative memory model is a neural network in which the
sequence of patterns is embedded as an attractor through Hebbian
(correlation) learning
\cite{Amari1988,During1998,Katayama2001,Kawamura2002}. When the number
of patterns, $p$, is of the order ${\cal O}(N)$, where $N$ is the number
of processing units, the model has frustrated interactions
\cite{Rieger1998}. The properties in stationary states were
analyzed exactly using the path-integral method
\cite{During1998,Katayama2001} because the theoretical treatment of the
transient was difficult. However, the transient of the model was
recently rigorously analyzed~\cite{Kawamura2002}.

The non-monotonicity of processing units (a larger absolute value of the
local field tends to make their state opposite that of the local
field) gives a system superior properties, e.g., enhanced storage
capacity, fewer spurious states, and a super retrieval phase
\cite{Morita1993,ShiinoFukai1993,Okada1996}. 
The systems with non-monotonic units have chaotic behaviors. Dynamic
theories are indispensable for analyzing the chaotic behaviors. The
dynamical mean-field theory \cite{Sompolinsky1988} is exact in the limit
of $N\to\infty$, and one can analyze the chaotic behaviors in random
neural networks. 
Only approximated theories, e.g., Gaussian approximation
\cite{Amari1988,Okada1996,AmariMaginu1988,NishimoriOpris1993,Okada1995,Kawamura1999}
or steady-state approximation \cite{During1998,Katayama2001}, have been
used to investigate the occurrence of these chaotic behaviors
in the associative memory models.

In our previous work \cite{Kawamura2003}, we constructed
bifurcation diagrams of a non-monotonic system. 
We showed chaotic behaviors in a non-monotonic sequential associative
memory model at absolute zero and demonstrated that the
chaos  occurs only when it has some degree of frustration. 

In this paper, we analyze bifurcations in our model at a finite
temperature.
We note that the microscopic behavior is stochastic while the
macroscopic one is deterministic \cite{Kawamura2002}.  We can therefore
analyze its macroscopic dynamics rigorously and construct two-parameter
bifurcation diagrams from our order-parameter equations. The structure
of the bifurcation is changed by the finite temperature effect. We
analytically show the area of a cusp point and the coexistence of
attractors, which do not appear at absolute zero.

\section{Sequential associative memory model}

Consider a sequential associative memory model consisting of
$N$ units or neurons. The state of the units takes $\sigma_i(t)=\pm1$
and is updated synchronously with probability
\begin{equation}
 \Prob\left[\sigma_i(t+1)\vert h_i(t)\right]
  =\frac12\left[1+\sigma_i(t+1)F\left(h_i(t)\right) \right] ,
  \label{eqn:dynamics}  
\end{equation}
\begin{equation}
 h_i(t)=\sum_{j=1}^NJ_{ij}\sigma_j(t)+I_i(t) ,
  \label{eqn:hi} 
\end{equation}
where $J_{ij}$ is the coupling, $I_i(t)$ is the threshold or external
input, and $h_i(t)$ is the local field. Function $F\left(h\right)$
is a non-monotonic function given by
\begin{equation}
 F\left(h\right)=\tanh\beta h-\tanh\beta\left(h-\theta\right)
  -\tanh\beta\left(h+\theta\right) ,
\end{equation}
where $\beta$ is the inverse temperature ($\beta=1/T$), and $\theta$ is
the non-monotonicity. When $T=0$, the update rule of the model is
deterministic:
\begin{equation}
 \sigma_i(t+1)=\sgn\left(h_i(t)\right)-\sgn\left(h_i(t)-\theta\right)-\sgn\left(h_i(t)+\theta\right) .
  \label{eqn:DetDynamics}
\end{equation}
When the absolute value of the local field is larger then $\theta$, the
sign of the state is opposite that of the local field, i.e., 
$\sigma_i(t+1)=-\sgn\left(h_i(t)\right)$. Coupling $J_{ij}$ stores
$p$ random patterns, $\vec{\xi}^{\mu}=(\xi^{\mu}_1,\cdots,\xi^{\mu}_N)^T$,
so as to retrieve the patterns sequentially:
$\vec{\xi}^0\to\vec{\xi}^1\to\cdots\vec{\xi}^{p-1}\to\vec{\xi}^0$.
It is given by
\begin{equation}
 J_{ij}=\frac1N\sum_{\mu=0}^{p-1}\xi^{\mu+1}_i\xi^{\mu}_j ,
  \label{eqn:Jij}
\end{equation}
where $\vec{\xi}^{p}=\vec{\xi}^0$. The number of stored patterns is
given by $p=\alpha N$, where $\alpha$ is the {\itshape loading
rate}. Each component of the patterns is assumed to be an independent
random variable that takes a value of either $+1$ or $-1$ based on
\begin{equation}
 \Prob\left[\xi_i^{\mu}=\pm1\right]=\frac{1}{2} .
\end{equation}
We determine the initial state, $\vec{\sigma}(0)$, based on
\begin{equation}
 \Prob[\sigma_i(0)=\pm1] = \frac{1\pm m(0) \xi^0_i}{2}. 
\end{equation}
The overlap, the direction cosine between $\vec{\sigma}(0)$ and 
$\vec{\xi}^0$, converges to $m(0)$ as $N\to\infty$.

\section{Macroscopic state equations}

To discuss the transient, we introduce macroscopic state equations by
using the path-integral method 
\cite{During1998,Katayama2001,Kawamura2002}. Generating function
$Z[\vec{\psi}]$ is defined as
\begin{eqnarray}
 Z[\vec{\psi}] &=& 
 \sum_{\vec{\sigma}(0),\cdots,\vec{\sigma}(t)} 
 p\left[\vec{\sigma}(0),\vec{\sigma}(1),\cdots,\vec{\sigma}(t)\right]
 %\nonumber \\  && \times
  \exp\left(-i\sum_{s<t}\vec{\sigma}(s)\cdot\vec{\psi}(s)\right) ,
 \label{eqn:Z0} 
\end{eqnarray}
where $\vec{\psi}=\left(\vec{\psi}(0),\cdots,\vec{\psi}(t-1)\right)$.
State $\vec{\sigma}(s)=(\sigma_1(s),\cdots,\sigma_N(s))^T$ denotes
the state of the spins at time $s$, and path probability
$p\left[\vec{\sigma}(0),\vec{\sigma}(1),\cdots,\vec{\sigma}(t)\right]$
denotes the probability of taking the path from initial state
$\vec{\sigma}(0)$ to state $\vec{\sigma}(t)$ at time $t$ through
$\vec{\sigma}(1),\vec{\sigma}(2),\cdots,\vec{\sigma}(t-1)$. Since the
dynamics, eq.~(\ref{eqn:dynamics}), is a Markov chain, the path probability
is given by
\begin{equation}
 p\left[\vec{\sigma}(0),\vec{\sigma}(1),\cdots,\vec{\sigma}(t)\right]
  =p\left[\vec{\sigma}(0)\right]\prod_{s<t}
  \prod_{i}\frac{1}{2}\left[1+\sigma_i(s+1)F\left(h_i(s)\right)\right] .
\end{equation}

The generating function involves the following order parameters:
\begin{eqnarray}
 m(s) &=& i\lim_{\vec{\psi}\to0}\frac{1}{N}\sum_{i=1}^N \xi_i^s
  \frac{\partial Z[\vec{\psi}]}{\partial \psi_i(s)} 
  \label{eqn:def_m} , \\
 G(s,s') &=& i\lim_{\vec{\psi}\to0}\frac{1}{N}\sum_{i=1}^N
  \frac{\partial^2 Z[\vec{\psi}]}{\partial\psi_i(s)\partial I_i(s')} , \\ 
 C(s,s') &=& -\lim_{\vec{\psi}\to0}\frac{1}{N}\sum_{i=1}^N
  \frac{\partial^2 Z[\vec{\psi}]}{\partial\psi_i(s)\partial\psi_i(s')} 
  \label{eqn:def_C} .
\end{eqnarray}
Order parameter $m(s)$ corresponds to the overlap, which represents the
direction cosine between state $\vec{\sigma}(s)$ and retrieval pattern
$\vec{\xi}^s$ at time $s$. $G(s,s')$ and $C(s,s')$ are the response and
correlation functions, respectively, between time $s$ and
$s'$. Therefore, the problem of evaluating the macroscopic dynamics
leads to the problem of evaluating the generating function.

We consider the case of thermodynamic limit $N\to\infty$ and
analyze $Z[\vec{\psi}]$ using the saddle point method. Since $N\to\infty$
and stored patterns $\vec{\xi}^{\mu}$ are random patterns, we can
assume self-averaging with respect to the realization of disorder;
that is, we would like to average $Z[\vec{\psi}]$ over the uncondensed
patterns. 
And then using the normalization condition $Z\left[0\right]=1$
\cite{DominicisPeliti1978,Dominicis1978}, we can
eliminate invalid order parameters and derive effective order parameters.
We can therefore obtain a rigorous solution using the path-integral
method \cite{Kawamura2002}.

Finally, we obtain the following macroscopic state equations from
$Z[\vec{\psi}]$ when $I_i(s)=0$:
\begin{eqnarray}
 m(s) &=& \left<\xi^{s}\int D_z F\left(\xi^{s}m(s-1)
				 % \right.\right.\nonumber \\ && \left.\left.
           +z\sqrt{\alpha R(s-1,s-1)}\right)\right>_{\xi} ,
 \label{eqn:mt_xi}
\end{eqnarray}
\begin{eqnarray}
 R(s,s')&=& C(s,s') 
  +G(s,s-1)G(s',s'-1)R(s-1,s'-1) , 
 \label{eqn:R}
\end{eqnarray}
\begin{eqnarray}
 G(s,s-1) &=& \frac{1}{\sqrt{\alpha R(s-1,s-1)}} \nonumber \\
 &\times& \left< \int \!D_zz F
   \left(\xi^{s}m(s-1)+z\sqrt{\alpha R(s-1,s-1)}\right)\! \right>_{\xi} ,
 \label{eqn:Gt_xi}
\end{eqnarray}
\begin{eqnarray}
 C(s,s')
 &=& \left<\int\frac{d\vec{z}}{2\pi\vert\vec{R}_{11}\vert^{\frac{1}{2}}}
      \exp\left[-\frac{1}{2}\vec{z}\cdot\vec{R}_{11}^{-1}\vec{z}\right]
	    F\left(\xi^{s}m(s-1)+\sqrt{\alpha}z(s-1)\right)
	  \right. \nonumber \\
 && \left. \times F\left(\xi^{s'}m(s'-1)+\sqrt{\alpha}z(s'-1)\right)
			 \right>_{\xi} ,
			 \label{eqn:Ct_xi} 
\end{eqnarray}
where $D_z=\frac{dz}{\sqrt{2\pi}}\exp[-\frac{1}{2}z^2]$, and
$\left<\cdot\right>_{\xi}$ denotes the average over all $\xi$'s. Matrix
$\vec{R}_{11}$ is a $2\times2$ matrix consisting of the elements of
$\vec{R}$ at times $s-1$ and $s'-1$, and $\vec{z}=[z(s-1),
z(s'-1)]^T$. From eqs.~(\ref{eqn:mt_xi})--(\ref{eqn:Ct_xi}), $C(s,s')=0$
and $R(s,s')=0$ when $s\neq s'$.  Since $G(s,s-1)$ and $C(s,s)$ can be
described using only $m(s-1)$ and $R(s-1,s-1)$, macroscopically this
system is a two-degree-of-freedom system of $m(s)$ and $R(s,s)$. 
Since we can easily calculate the Gaussian integrals, 
we can analyze the transient dynamics exactly even if the
network fails in retrieval.
We note that $m(s)$ is an odd function and $R(s,s)$ is an even function,
since the function $F(h)$ is an odd function. Therefore, the map by the
macroscopic state equations is line symmetric with respect to the line
$m=0$.

Besides these dynamic macroscopic state equations, the fixed points of
the system are required in order to analyze the bifurcation of the
system. We set $m(t)\to m, G(t,t-1)\to G$ and $ R(t,t)\to r$ when
$t\to\infty$. Then, the previously obtained stationary state equations 
\cite{During1998,Katayama2001} are re-derived using our dynamic theory:
\begin{eqnarray}
 m &=& \left<\xi\int D_z F
        \left[\xi m+z\sqrt{\alpha r}\right] \right>_{\xi} ,
  \label{eqn:stat_m} \\
 G &=& \frac{1}{\sqrt{\alpha r}}\left<\int D_zz F
        \left[\xi m+z\sqrt{\alpha r}\right] \right>_{\xi} , \\
 r &=& \frac{1}{1-G^2} .
  \label{eqn:stat_r}
\end{eqnarray}

\section{Macroscopic dynamics}

\begin{figure}[tb]
 \hfill
 (a) \includegraphics[width=60mm]{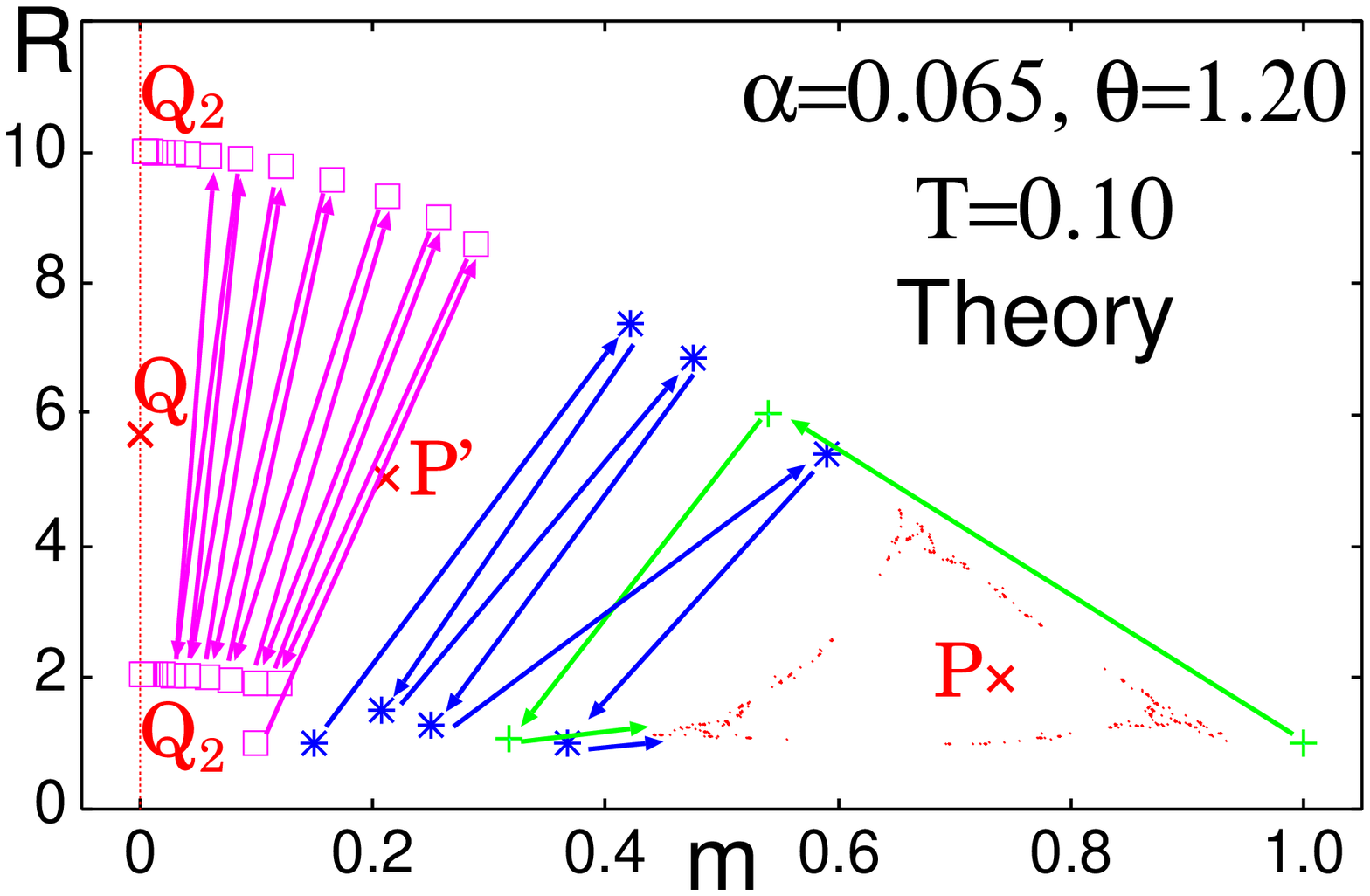}
 \hfill
 (b) \includegraphics[width=60mm]{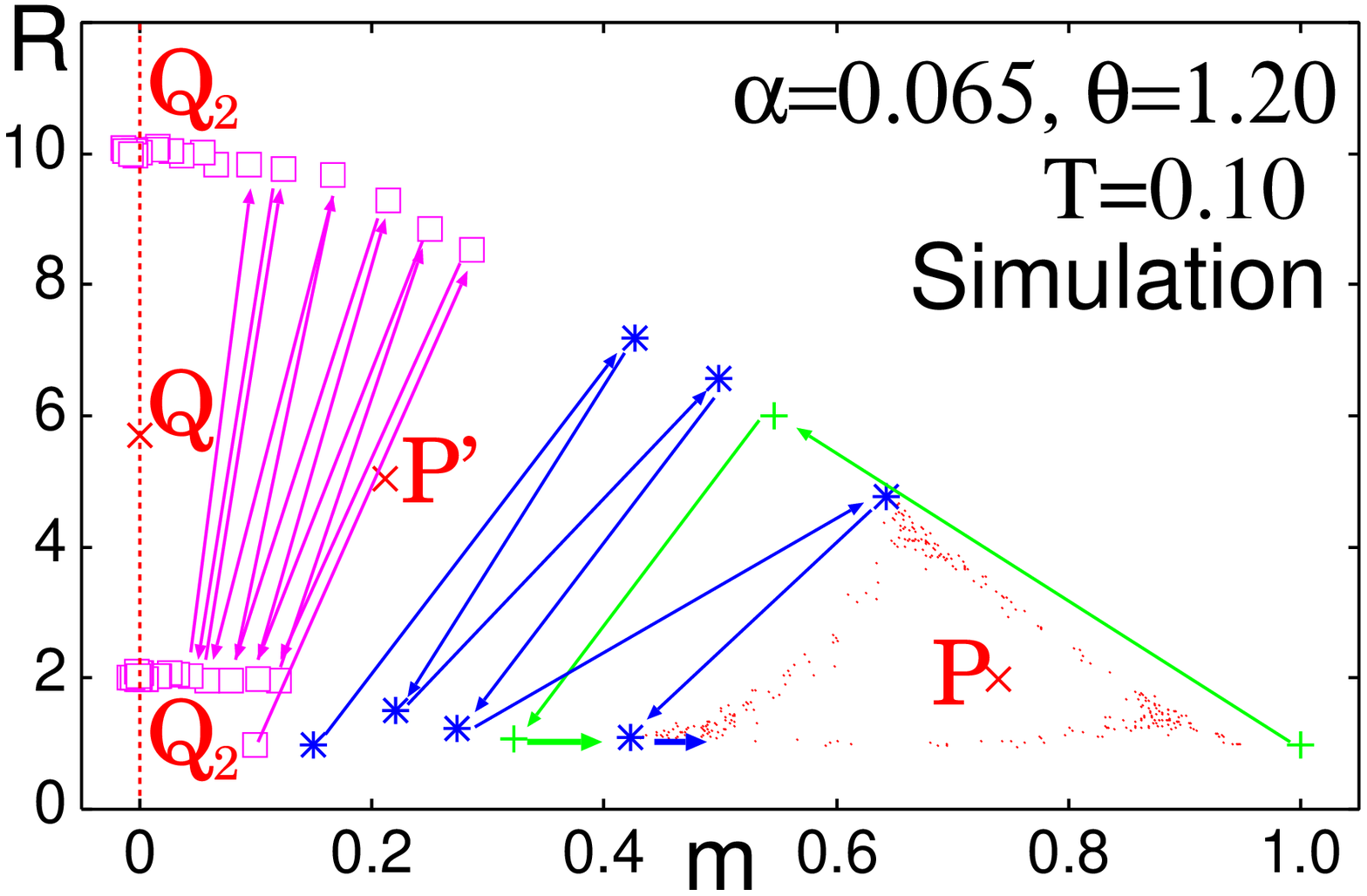}
 \hfill\mbox{}
 \caption{Transition of overlap $m(t)$ and variance of crosstalk noise
 $\alpha R(t,t)$ for $\alpha=0.065, \theta=1.20, T=0.10$. 
 Fixed points $P,P',Q$ are from stationary state equations:
 (a) our theory, (b) simulation ($N=100,000$).
 }
 \label{fig:mr}
\end{figure}

We can obtain the macroscopic dynamics
(\ref{eqn:mt_xi})--(\ref{eqn:Ct_xi}) from the stochastic microscopic
dynamics. Moreover, the Jacobian matrix, $\vec{J}(s)$, can be easily
calculated from these equations:
\begin{equation}
 \vec{J}(s) = \left[\begin{array}{cc}
	 \frac{\partial m(s)}{\partial m(s-1)} & \frac{\partial m(s)}{\partial R(s-1,s-1)} \\
	       \frac{\partial R(s,s)}{\partial m(s-1)} & \frac{\partial R(s,s)}{\partial R(s-1,s-1)} 
	      \end{array}\right] .
\end{equation}
We can therefore classify fixed points according to their eigenvalues,
$\lambda$.
We analyze the transient for a finite temperature, e.g.,
$T=0.10$.  Figure~\ref{fig:mr} shows the transition of the overlap,
$m(t)$, and the variance of the crosstalk noise, $\alpha R(t,t)$. The
graphs show the results obtained using (a) our theory and (b) computer
simulation with $N=100,000$, where loading rate $\alpha$ is $0.065$
and non-monotonicity $\theta$ is $1.20$. The cross marks ($P,P',Q$)
are fixed points. Point $P$ is a repellor (unstable focus), since the
eigenvalues are $\lambda=-0.21\pm1.40i$; 
$P'$ is a repellor (unstable node) because $\lambda=-1.30, 1.18$; and
$Q$ on line $m=0$ is a saddle node (orientation reversing) because 
$\lambda=-1.29, 0.91$.
There is a period-$2$ attractor, $Q_2$, that attracts
the trajectories with initial state $m(0)\approx 0$. Moreover, there is
a chaotic attractor around repellor $P$. The results obtained using our
theory agree with those using computer simulation.
Since our dynamic macroscopic state equations are derived exactly,
the difference between the theoretical analysis and the computer
simulation is due to finite size effect.

\section{Bifurcation diagram for $T=0$}

\begin{figure}[tb]
 \begin{center}
   (a) \includegraphics[width=100mm]{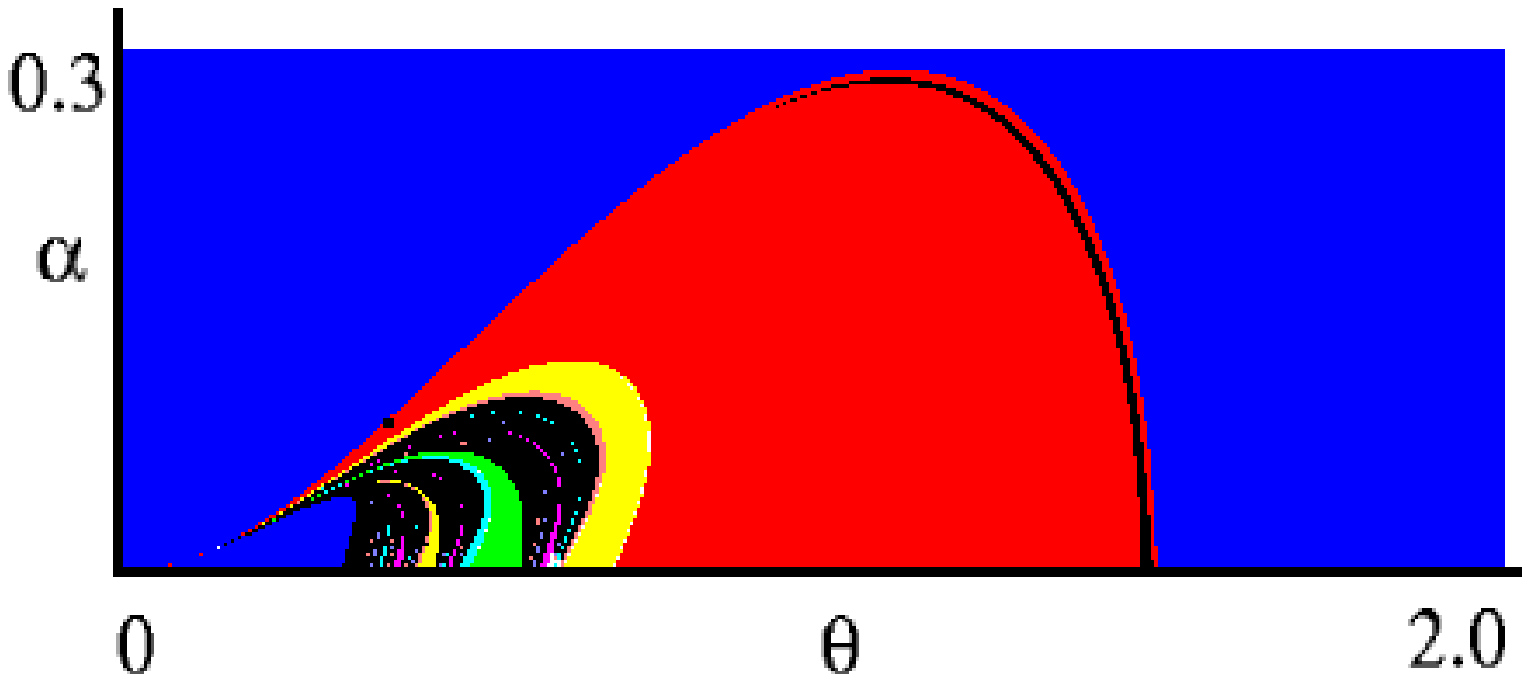}

   (b) \includegraphics[width=100mm]{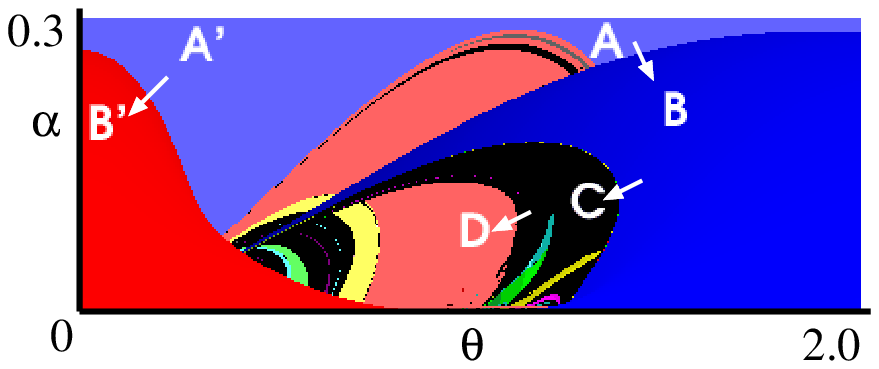}
 \end{center} 
 \caption{Two-parameter bifurcation diagram ($\theta,\alpha$) for (a) 
 fixed point $Q$ and (b) fixed point $P$ at $T=0$. 
 Blue region represents period-$1$ attractors, red period-$2$,
 green period-$3$, yellow period-$4$, purple period-$5$, sky
 blue period-$6$, and black for more than six periods, quasi-periodic
 or chaotic.
 }
 \label{fig:phase}
\end{figure}

We investigated the relationships of the invariant sets shown in
Fig.~\ref{fig:mr} in two-parameter space with respect
to $(\theta,\alpha)$. Line $m=0$ is an invariant set of the
macroscopic state equations, and the map of the system is line symmetric
with respect to invariant line $m=0$, as stated above. 
The dynamic structure on this invariant line obeys a one-dimensional map
with respect to $R(t,t)$ \cite{Kawamura2003}.
Figure~\ref{fig:phase} (a) shows a two-parameter bifurcation diagram for
the attractor on invariant line $m=0$, and (b) shows one for the
attractor around repellor $P$ for $T=0$. The blue region
represents the period-$1$ attractors, red 
period-$2$, green period-$3$, yellow period-$4$, purple period-$5$, sky
blue period-$6$, and black for more than six periods, quasi-periodic or
chaotic.  In Fig. (a), as $\theta$ decreases, a period-$1$ attractor,
$Q$, bifurcates to a period-$2$ attractor, $Q_2$, and evolves into a
chaotic attractor due to the period-doubling cascade.
In Fig. (b), some regions are denoted by $A,A',B,B',C,$ and $D$,
and we can find bifurcations on the boundaries between these regions.

\subsection{Transient}

\begin{figure}[tb]
 \begin{center}
  \hfill (a) \includegraphics[width=62mm]{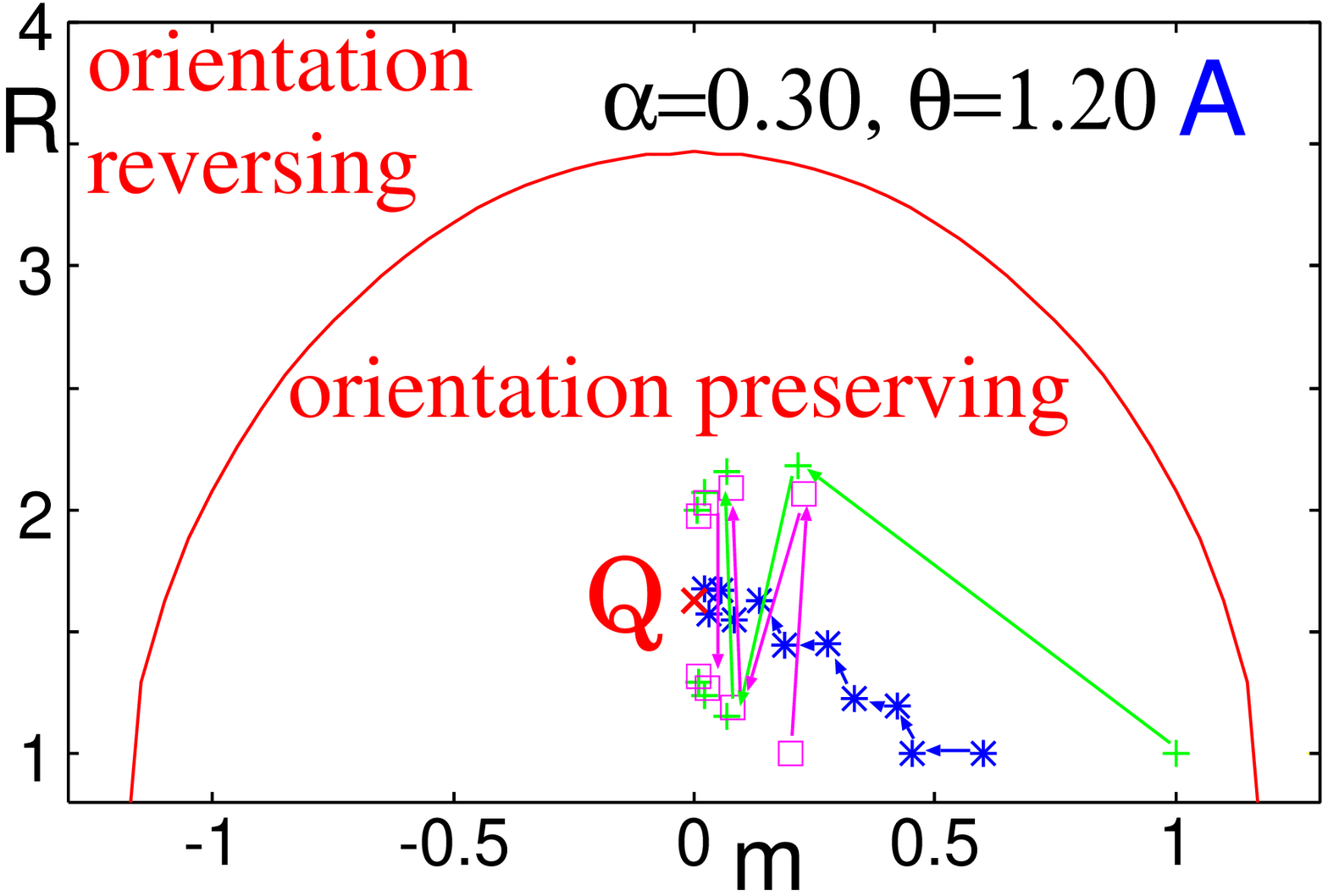}
  \hfill (b) \includegraphics[width=62mm]{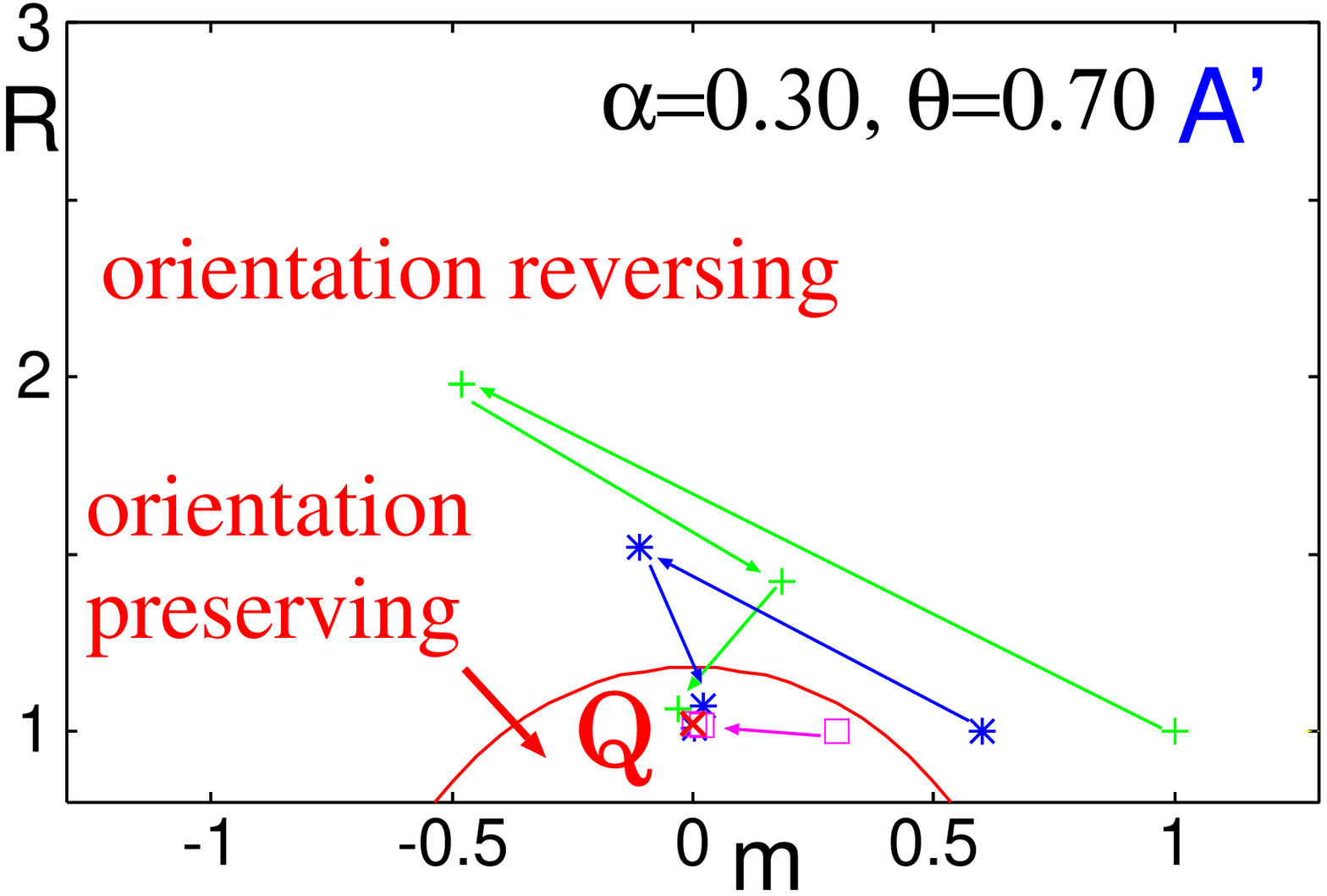}
  \hfill \mbox{}
 \end{center} 
 \caption{Attractor $Q$ in (a) region $A$ ($\alpha=0.30, \theta=1.20$)
 and (b) region $A'$ ($\alpha=0.30, \theta=0.70$) at $T=0$ by our
 theory.
 Area inside semielliptic arc is orientation-preserving area, and
 area outside is orientation-reversing area.}
 \label{fig:mr_dynamicsA}
\end{figure}

Figure~\ref{fig:mr_dynamicsA} shows the transient in regions $A$ and
$A'$. There is only a period-$1$ attractor, $Q$, on the invariant
line. Since the map by our macroscopic state equations is irreversible,
there is an orientation-preserving area inside the semielliptical arc,
and an orientation-reversing area outside the arc. The
orientation-preserving area shrinks as $\theta$ decreases.
Therefore, the transient in $A$ differs from that in $A'$.  In both
cases, since the stored patterns are unstable, the associative
memory fails to retrieve one from any initial state.

\begin{figure}[htb]
 \begin{center}
  \hfill (a)\includegraphics[width=62mm]{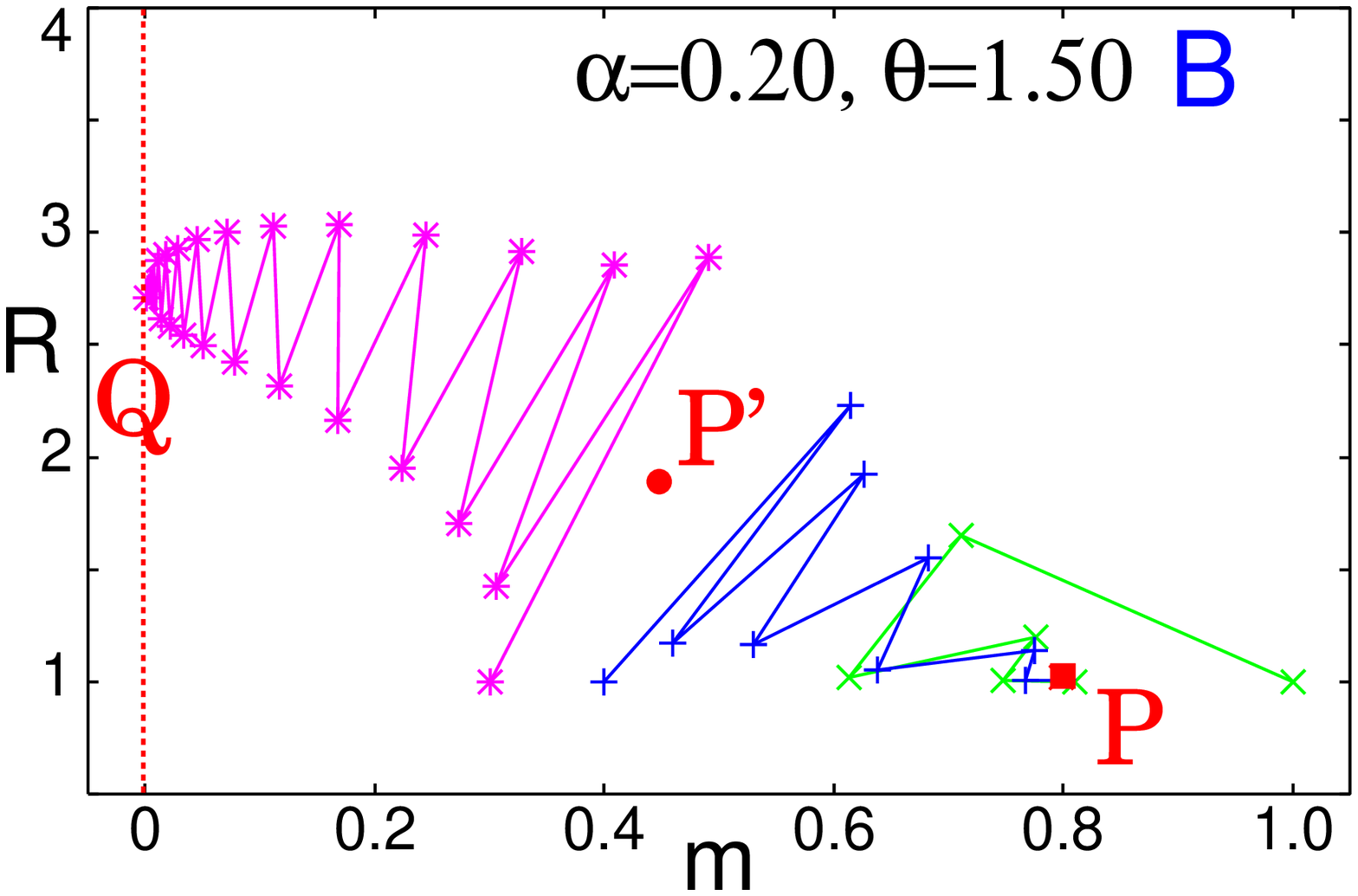}
  \hfill (b)\includegraphics[width=62mm]{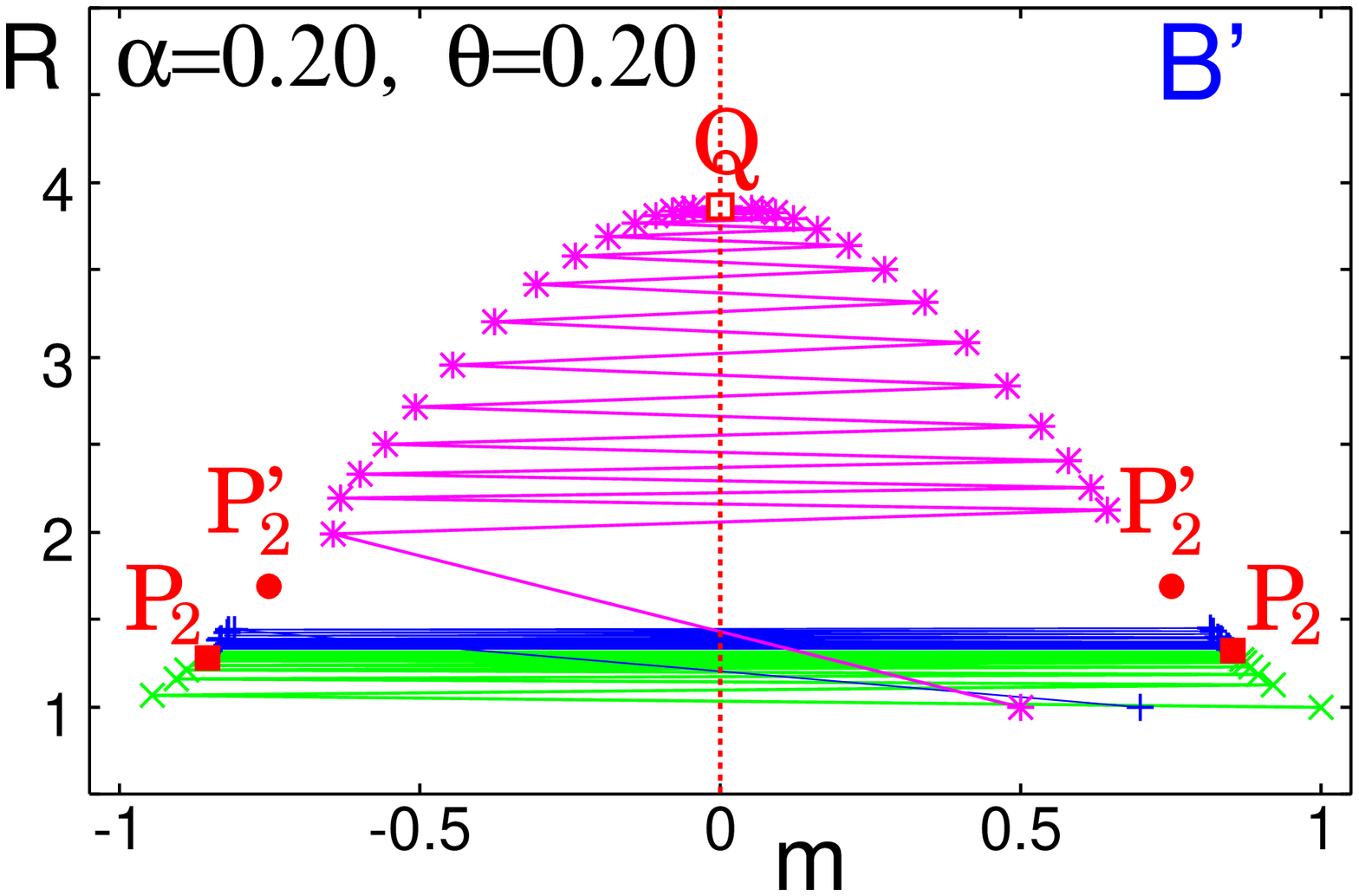}
  \hfill \mbox{}
 \end{center} 
 \caption{Attractors $P$ and $P_2$ in
 (a) region $B$ ($\alpha=0.20, \theta=1.50$) and (b) region
 $B'$ ($\alpha=0.20, \theta=0.20$) at $T=0$ by our theory.}
 \label{fig:mr_dynamicsB}
\end{figure}

Figure~\ref{fig:mr_dynamicsB} shows the transient in regions $B$ and
$B'$. In region $B$, there is both a period-$1$ attractor, $Q$, on the
invariant line and a period-$1$ attractor, $P$, near $m=1$. The
orientation-reversing area is far from the origin. In this case, since
the stored patterns are stable, the associative memory can retrieve one
when the state is in the basin of the attraction of $P$. Additionally,
in region $B'$, there is both attractor $Q$ and period-$2$ attractor
$P_2$.  The orientation-preserving area shrinks to near the origin.
Attractor $P_2$ is a sign-reversing state near line $m=\pm1$.  In this
case, the stored patterns are unstable, and the memory retrieves the
stored pattern and its reverse one in turn when the state is in the
basin of attraction of $P_2$.

\begin{figure}[htb]
 \begin{center}
  \hfill (a)\includegraphics[width=62mm]{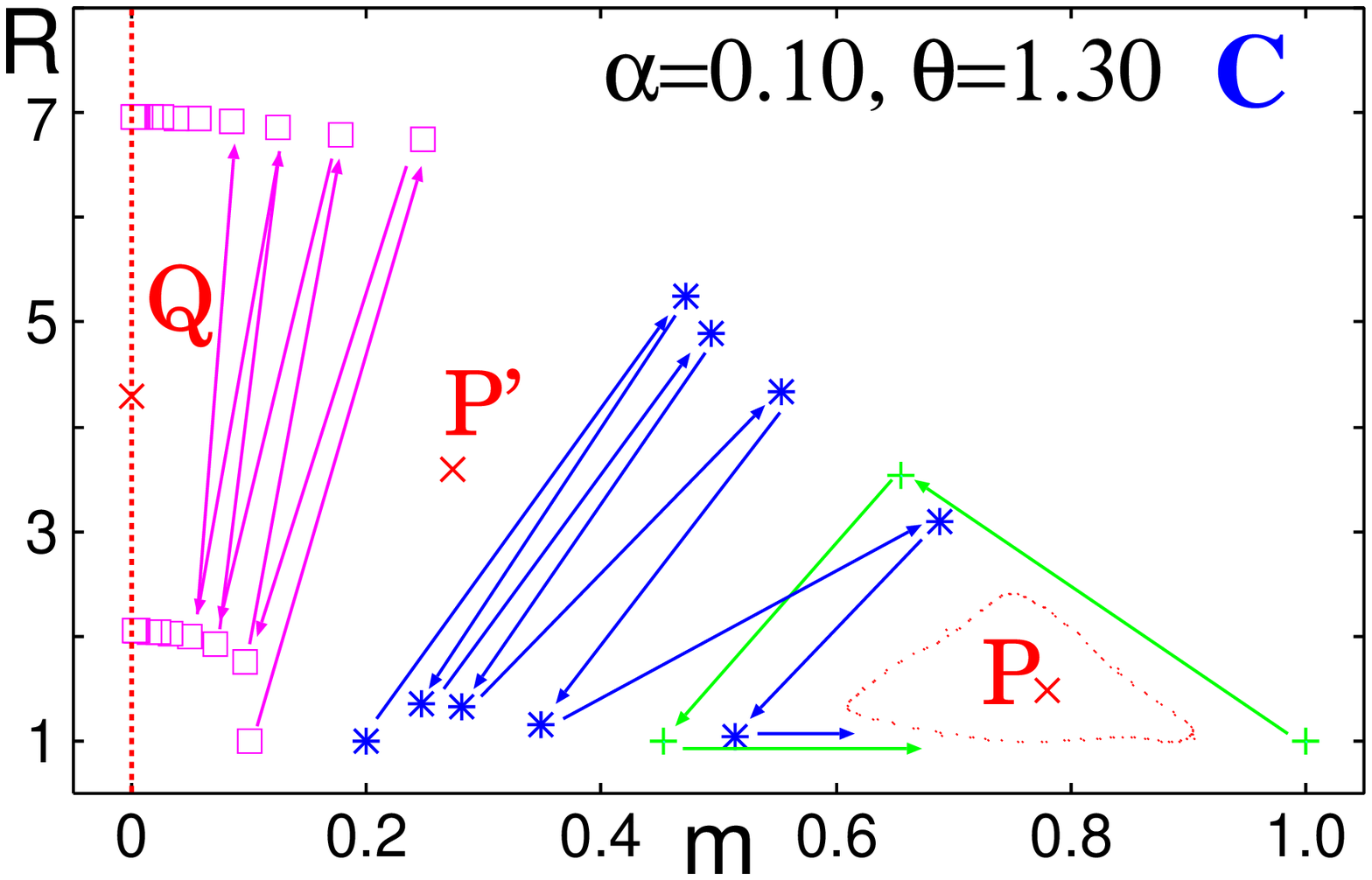}
  \hfill (b)\includegraphics[width=62mm]{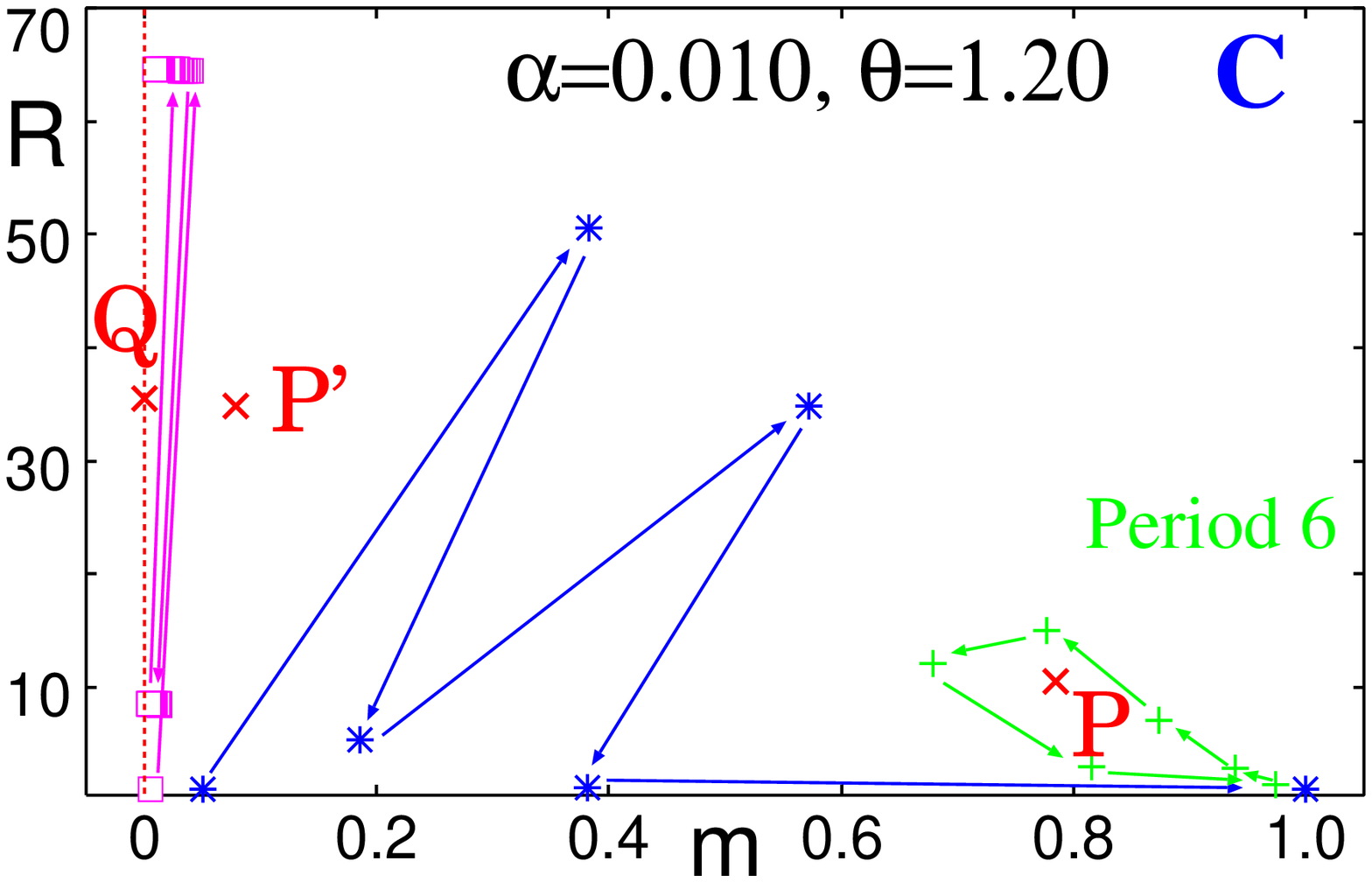}
  \hfill \mbox{}

  \hfill (c)\includegraphics[width=62mm]{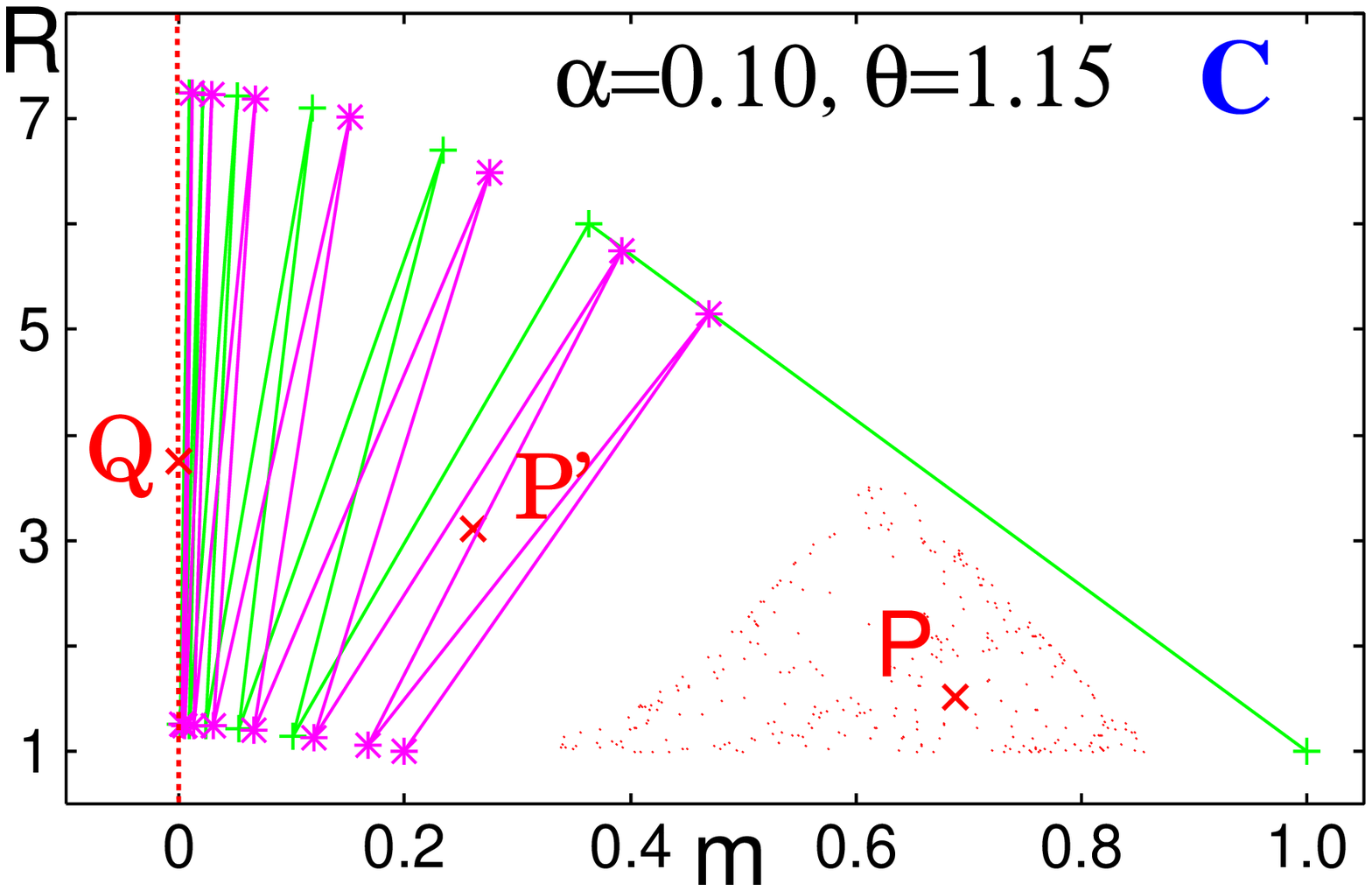}
  \hfill \mbox{}
 \end{center} 
 \caption{Attractors around $P$ in region $C$: (a) quasi-periodic
 attractor ($\alpha=0.10, \theta=1.30$), 
 (b) period-$6$ attractor ($\alpha=0.01, \theta=1.20$), and 
 (c) chaotic attractor ($\alpha=0.10, \theta=1.15$).}
 \label{fig:mr_dynamicsC}
\end{figure}

Figure~\ref{fig:mr_dynamicsC} shows the transient in region $C$. There
is both a period-$2$ attractor, $Q_2$, on the invariant line and a
certain attractor near $m=1$. The attractor around repellor $P$ is
periodic, quasi-periodic, or chaotic. In this case, although the stored
patterns are unstable, there is a quasi-periodic or chaotic
attractor. The state, therefore, goes to this attractor instead of the
memory state. Since the overlap is non-zero, the associative memory
neither completely succeeds nor fails to retrieve patterns.

\begin{figure}[htb]
 \begin{center}
  \includegraphics[width=62mm]{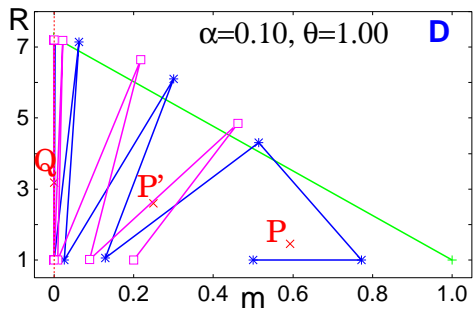}
 \end{center} 
 \caption{Attractors around $P$ in region $D$ ($\alpha=0.10,
 \theta=1.00$) vanish due to boundary crisis. }
 \label{fig:mr_dynamicsD}
\end{figure}

Figure~\ref{fig:mr_dynamicsD} shows the transient in region $D$. There
is a period-$1$ attractor, $Q$, and a period-$2$ attractor, $Q_2$, on
the invariant line. In this case, since the stored patterns are 
unstable, the associative memory fails to retrieve one from any initial
state.

\subsection{Bifurcations}

The coexistence, as stated above, can be explained by the occurrence of
characteristic bifurcations on the boundary between regions. On
boundary $A\to B$, saddle node $P'$ and period-$1$ attractor $P$ are
generated by the saddle node bifurcation, leading to the existence of
both $Q$ and $P$. They are separated by the basin boundary constituted
by $P'$. 
The boundary $B\to A$ represents the storage capacity, i.e., the
critical loading rate.
On boundary $A'\to B'$, similarly, period-$2$ saddle node
$P_2'$ and period-$2$ attractor $P_2$ are generated by the saddle node
bifurcation, leading to the existence of both $Q$ and $P_2$. In
contrast, on boundary $B\to C$, period-$1$ attractor $P$ evolves into a
repellor due to the Hopf bifurcation, and a quasi-periodic attractor is
generated around the repellor. This attractor is sometimes phase-locked,
and then it evolves into a more complex quasi-periodic attractor
by the Hopf bifurcation again. The repellor inside the
quasi-periodic attractor then evolves into a {\itshape snap-back
repellor} \cite{Marotte1978}, and 
belt-like chaos appears. Finally, the chaos spreads and becomes a thick
chaotic attractor, including repellor $P$. On boundary $C\to D$, the
chaotic attractor disappears due to a {\itshape boundary crisis}
\cite{Grebogi1982} because it comes into contact with the basin boundary
constituted by $P'$. Therefore, in region $D$, there is only period-$2$
attractor $Q_2$ on the invariant line.

\section{Bifurcation diagram for $T>0$}

\begin{figure*}[tb]
 \hfill \includegraphics[width=75mm]{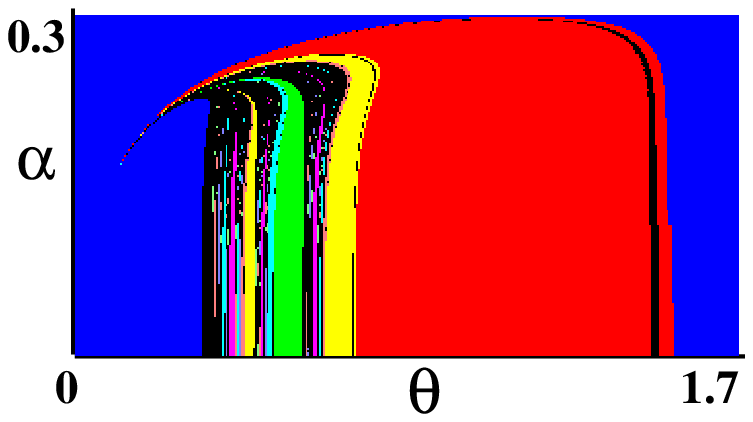}
 \hfill \includegraphics[width=75mm]{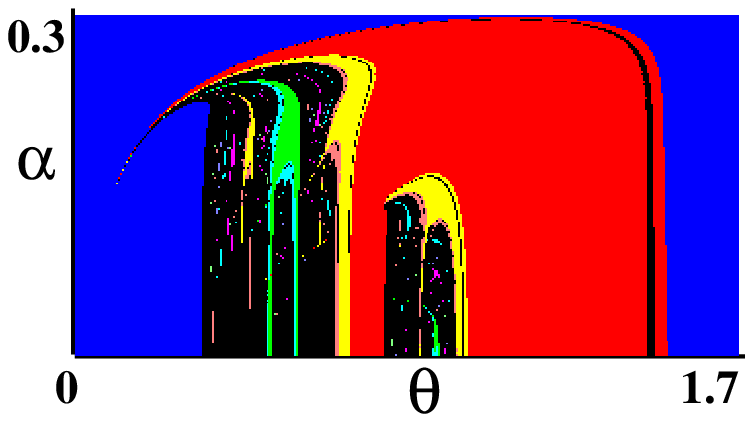}
 \hfill \mbox{} \\ 
 \vspace*{-1.5em}
 \hfill(a) $T=0$ \hfill(b) $T=0.05$\hfill \mbox{}
 %\vspace*{1em}

 \hfill \includegraphics[width=75mm]{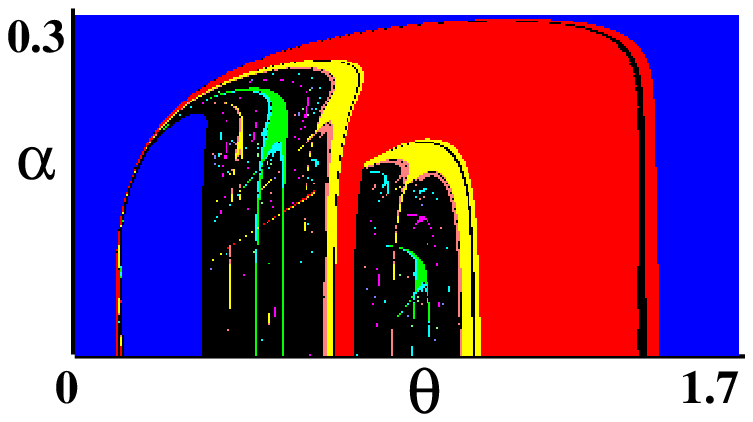}
 \hfill \includegraphics[width=75mm]{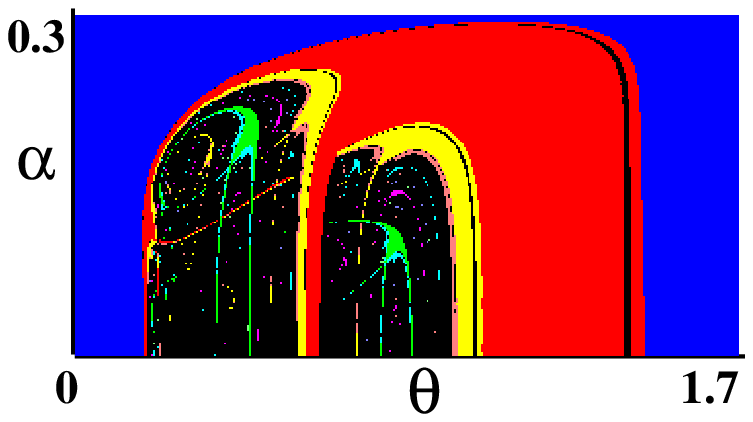}
 \hfill \mbox{} \\ 
 \vspace*{-1.5em}
 \hfill(c) $T=0.10$ \hfill(d) $T=0.15$\hfill \mbox{}
 %\vspace*{1em}

 \hfill \includegraphics[width=75mm]{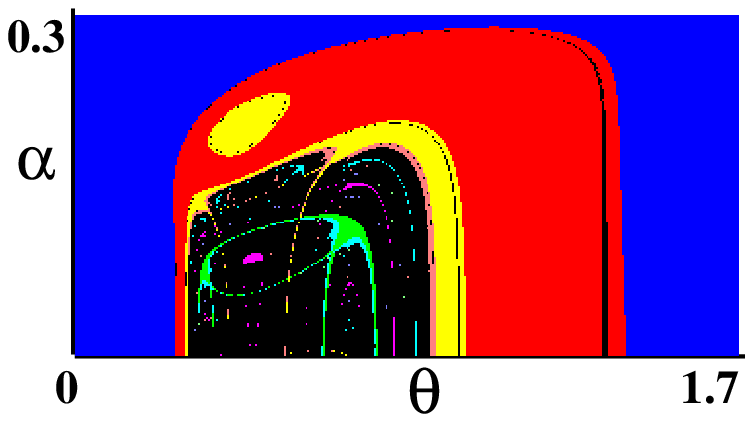}
 \hfill \includegraphics[width=75mm]{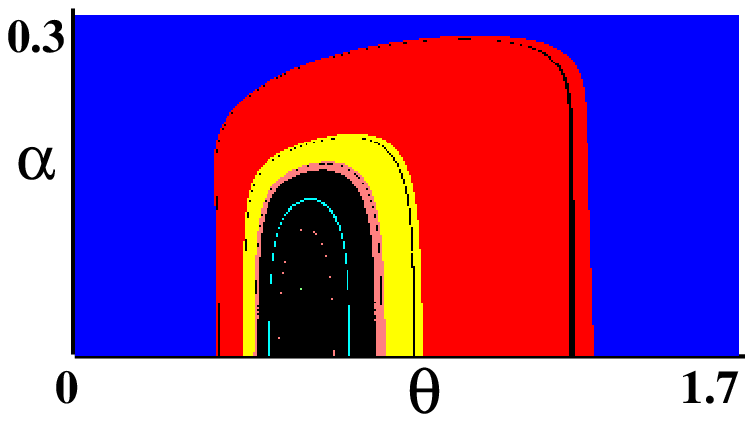}
 \hfill \mbox{} \\
 \vspace*{-1.5em}
 \hfill(e) $T=0.20$ \hfill(f) $T=0.25$\hfill \mbox{}
 %\vspace*{1em}

 \hfill \includegraphics[width=75mm]{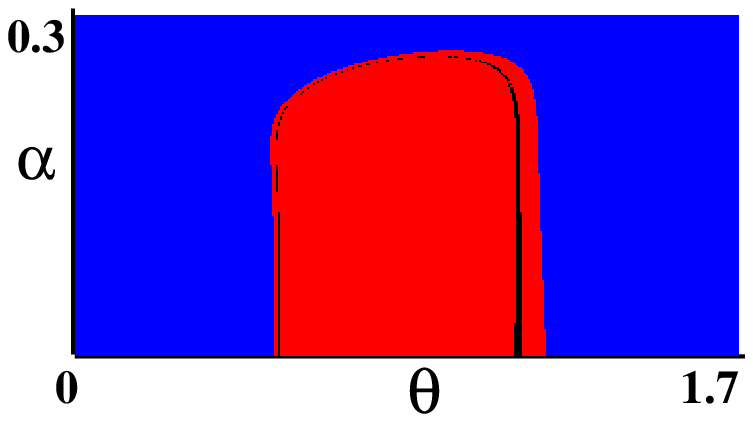}
 \hfill \includegraphics[width=75mm]{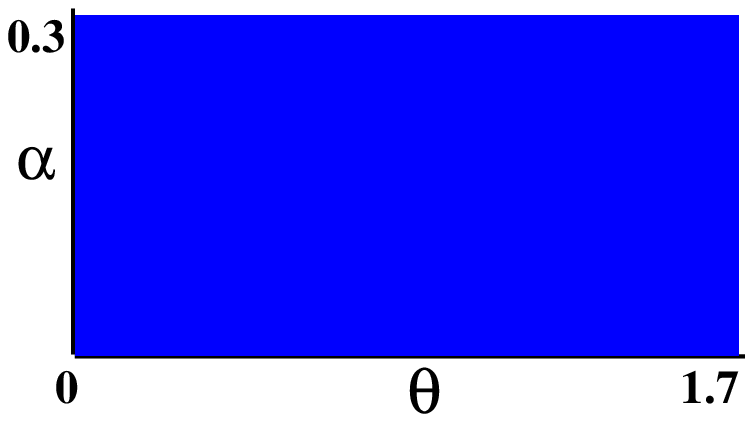}
 \hfill \mbox{} \\
 \vspace*{-1.5em}
 \hfill(g) $T=0.30$ \hfill(h) $T=0.35$\hfill \mbox{}

 \caption{Two-parameter bifurcation diagrams on invariant line
 $m=0$ for (a) absolute zero ($T=0$) and
 (b)--(h) finite temperature ($T=0.05-0.35$). 
 Abscissa denotes $\theta$ ($0<\theta<1.7$), and ordinate
 denotes $\alpha$ ($0<\alpha<0.302$) on logarithmic scale. 
 Colors denote period of attractors as in Fig.~\ref{fig:phase}.}
 \label{fig:diagramQ}
\end{figure*}

We constructed two-parameter bifurcation diagrams for several finite
temperatures. Figure~\ref{fig:diagramQ} shows the diagrams for an
attractor on invariant line $m=0$ for $T=0.0-0.35$. The
abscissa denotes $\theta$ ($0<\theta<1.7$), and the ordinate denotes
$\alpha$ ($0<\alpha<0.302$) on a logarithmic scale.  In the center of
the diagrams, we can find another bistable region in which 
attractors coexist.  As the temperature is increased, the bistable region
becomes larger.  For $T>0$, there is a {\itshape
fishhook} structure where the region of each periodic attractor divides 
into two regions.  When $T\geq 0.30$, the more-than-two-period
attractors and chaotic attractors disappear.

\begin{figure*}[tb]
 \hfill \includegraphics[width=75mm]{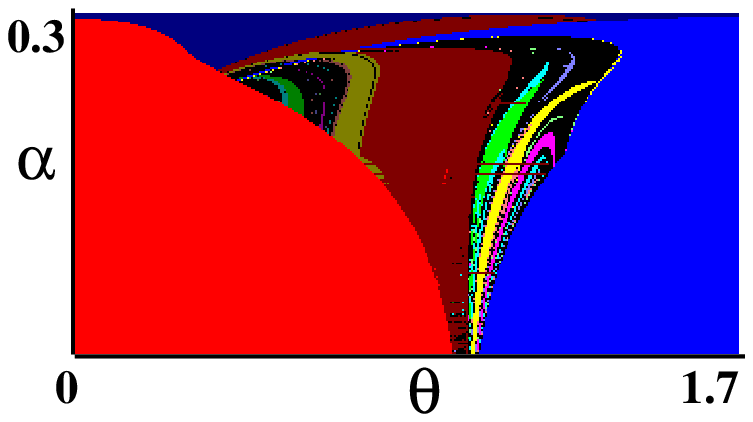}
 \hfill \includegraphics[width=75mm]{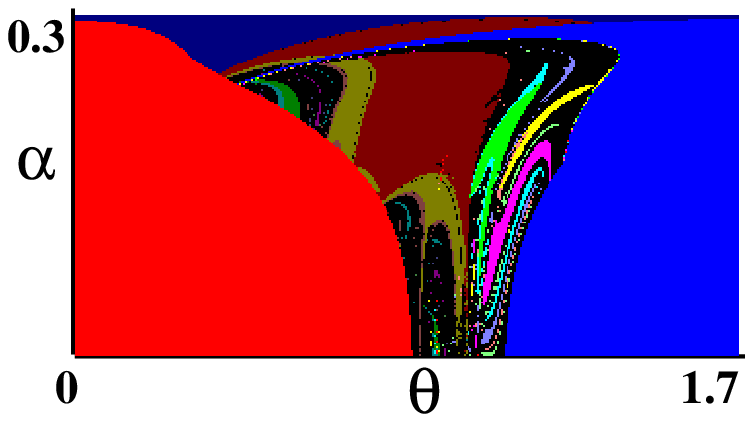}
 \hfill \mbox{} \\ 
 \vspace*{-1.5em}
 \hfill(a) $T=0$ \hfill(b) $T=0.05$\hfill \mbox{}
 %\vspace*{.5em}

 \hfill \includegraphics[width=75mm]{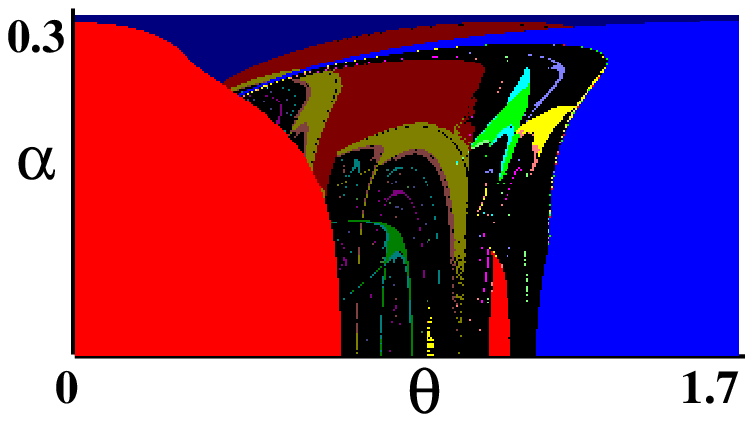}
 \hfill \includegraphics[width=75mm]{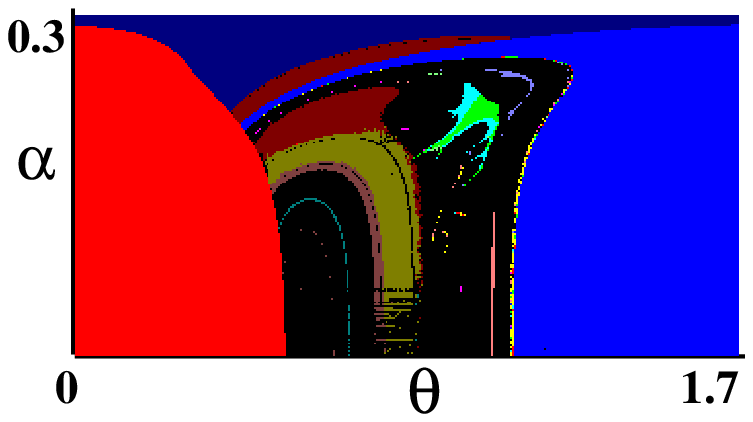}
 \hfill \mbox{} \\ 
 \vspace*{-1.5em}
 \hfill(c) $T=0.15$ \hfill(d) $T=0.25$\hfill \mbox{}
 %\vspace*{.5em}

 %%\hfill \includegraphics[width=68mm]{bifPall_0.10_log2.eps}
 %%\hfill \includegraphics[width=68mm]{bifPall_0.20_log2.eps}
 %%\hfill \mbox{} \\ 
 %%\hfill(e) $T=0.10$ \hfill(f) $T=0.20$\hfill \mbox{}
 %%\vspace*{.5em}

 \hfill \includegraphics[width=75mm]{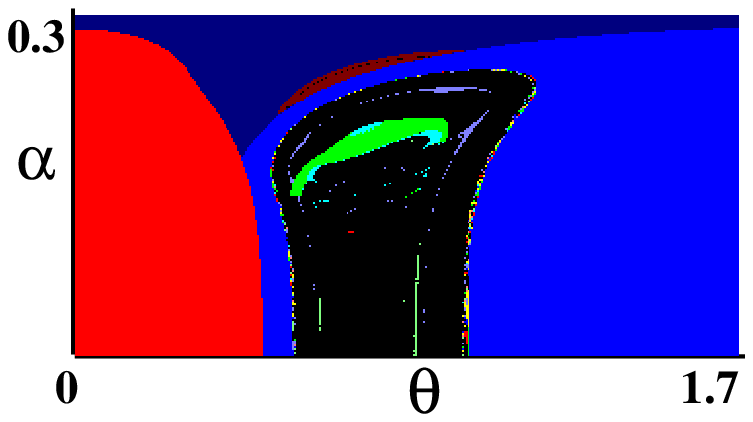}
 \hfill \includegraphics[width=75mm]{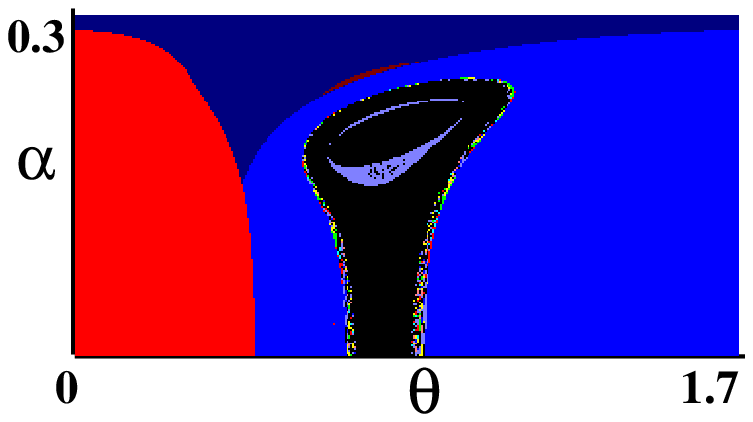}
 \hfill \mbox{} \\ 
 \vspace*{-1.5em}
 \hfill(e) $T=0.30$ \hfill(f) $T=0.32$\hfill \mbox{}
 %\vspace*{.5em}

 \hfill \includegraphics[width=75mm]{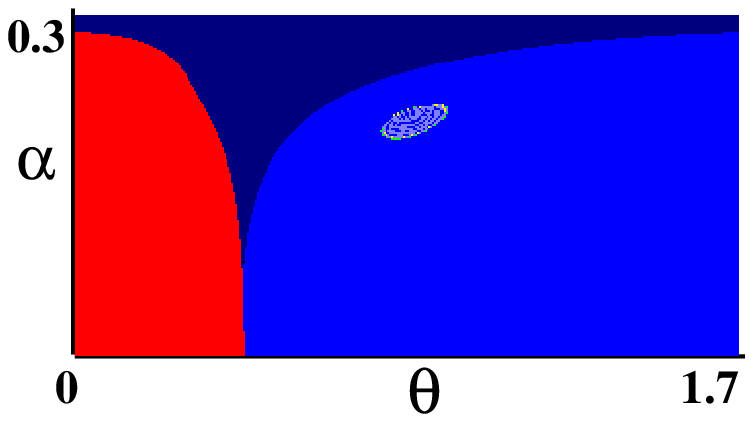}
 \hfill \includegraphics[width=75mm]{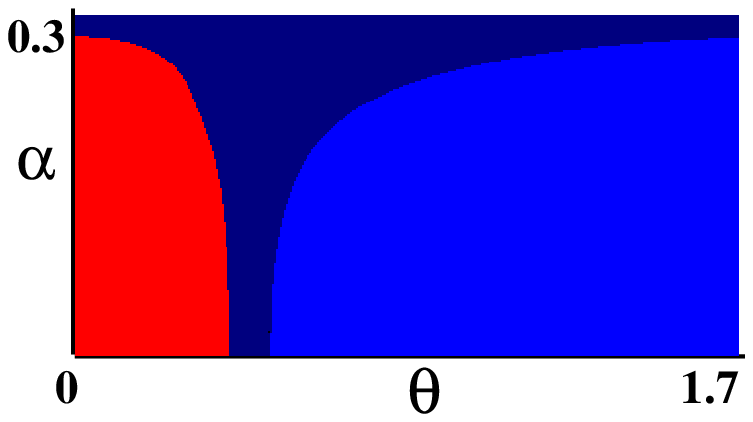}
 \hfill \mbox{} \\ 
 \vspace*{-1.5em}
 \hfill(g) $T=0.35$ \hfill(h) $T=0.40$\hfill \mbox{}

 \caption{Two-parameter bifurcation diagrams around fixed point $P$ 
 for (a) absolute zero ($T=0$) and (b)--(h) finite temperature
 ($T=0.05-0.40$).  
 Abscissa denotes $\theta$ ($0<\theta<1.7$), and ordinate
 denotes $\alpha$ ($0<\alpha<0.302$) on logarithmic scale.
 Colors denote period of attractors as in Fig.~\ref{fig:phase}.}
 \label{fig:diagramP}
\end{figure*}

Figure~\ref{fig:diagramP} shows the two-parameter bifurcation diagrams
for fixed point $P$ or of the attractors around $P$ for $T=0.0-0.40$.
The diagrams overlay those in Fig.~\ref{fig:diagramQ} since one can
see the coexistence of attractors $P$ and $Q$.
As the temperature is increased, the Hopf bifurcation set becomes an isolated
circle and disappears via the codimension-$2$ bifurcation set.
Since a fishhook structure is evident in the diagram, there is a cusp
point, i.e., a codimension-$2$ bifurcation set, which generates
a pair of saddle node bifurcation sets.

\begin{figure}[bt]
 \begin{center}
  \includegraphics[width=85mm]{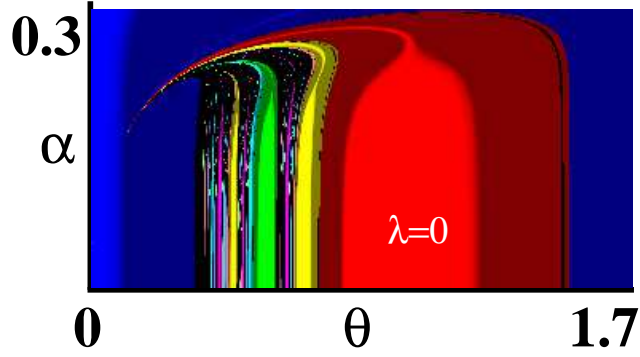}

  (a) $T=0$ 

  \includegraphics[width=85mm]{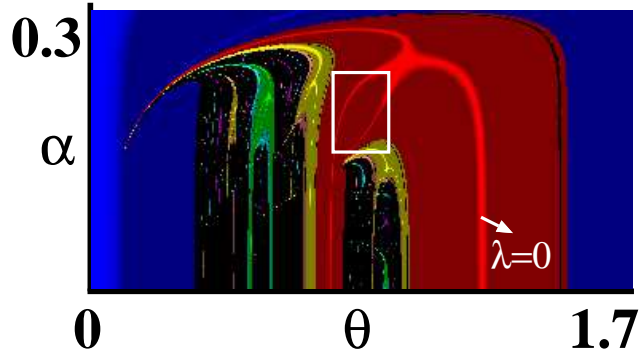}

  (b) $T=0.05$  
 \end{center}
 \caption{Curve of eigenvalue $\lambda=0$ and area of cusp point for
 (a) $T=0$ and (b) $T=0.05$. Cusp point is in rectangular area. }
 \label{fig:cusp}
\end{figure}

We show the region where the cusp point exists. Figure~\ref{fig:cusp}
shows the two-parameter bifurcation diagrams for fixed point $Q$,
which are graded by eigenvalue $\lambda$.  The bright lines denote
eigenvalue $\lambda=0$.  There is only
one curve for $\lambda=0$ at $T=0$, whereas there are two for
$T=0.05$. A cusp point is in the region surrounded by the curves.

\section{Discussion}

\begin{figure}[ht]
 \begin{center}
  \includegraphics[width=80mm]{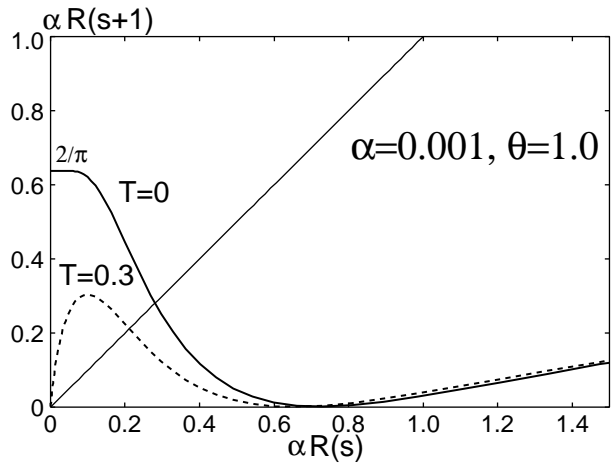}
 \end{center}
 \caption{Return map of variance of crosstalk noise, $\alpha R(t,t)$ for
 $\alpha=0.001, \theta=1.0$. 
 Solid and broken lines denote cases of $T=0$ and $T=0.3$, respectively.}
 \label{fig:RetMap}
\end{figure}

We can see that the chaotic region may be expanding to
$\alpha\to0$ in Figs. \ref{fig:diagramQ} and \ref{fig:diagramP}. 
We first discuss the cases of $\alpha=0$ and $\alpha\to0$. 
 Only one order-parameter, $m(t)$, dominates the
macroscopic behaviors of the present system without frustration of
interaction, that is, when the number of stored patterns is finite
($\alpha=0$).  We can easily show that there is no chaotic attractor in
this case.  A chaotic attractor appears when the system has
frustration, i.e., $\alpha\neq0$.  While the local field, $h_i(t)$,
obeys a $\delta$-function distribution when $\alpha=0$, it obeys a
Gaussian distribution with variance $\alpha R(t,t)$ when $\alpha \neq
0$.  
We show the simple case of $T=0$ and $m(t)=0$. 
From eqs.~(\ref{eqn:DetDynamics}), (\ref{eqn:R}), and (\ref{eqn:Gt_xi}), 
the variance of crosstalk noise, $\alpha R(t+1,t+1)$, can be given by
\begin{eqnarray}
 \alpha R(t+1,t+1) &=& \alpha +\frac{2}{\pi}
  \left\{1-2\exp\left(-\frac{\theta^2}{2\alpha R(t,t)}\right)\right\}^2. 
\end{eqnarray}
When $\alpha\to0$, $\alpha R(t+1,t+1)$ converges to 
\begin{equation}
 \alpha R(t+1,t+1) = \frac{2}{\pi}
  \left\{1-2\exp\left(-\frac{\theta^2}{2\alpha R(t,t)}\right)\right\}^2,
\end{equation}
since $R(t,t)$ takes a large value in inverse proportion to
$\alpha$. That is, $\alpha R(t+1,t+1)$ still takes a finite value even if
$\alpha\to0$. 
Figure~\ref{fig:RetMap} shows the return map of $\alpha R(t,t)$ for
$T=0$ and $0.3$. 
We can see that the fixed point is finite ($\alpha R(t,t)>0$). 
Because of this finite variance of the local field distribution, the
processing units, whose absolute values of local field $h_i(t-1)$ are
around non-monotonicity $\theta$, take different values, similar to
those in Bakers' Map.  This is the reason for the occurrence of
chaos in this model with frustration.  Non-trivial findings are that
chaos also appears when $\alpha\to0$ and that the phase is completely
different from when $\alpha=0$.  

\begin{figure}[tb]
 \begin{center}
  \includegraphics[width=80mm]{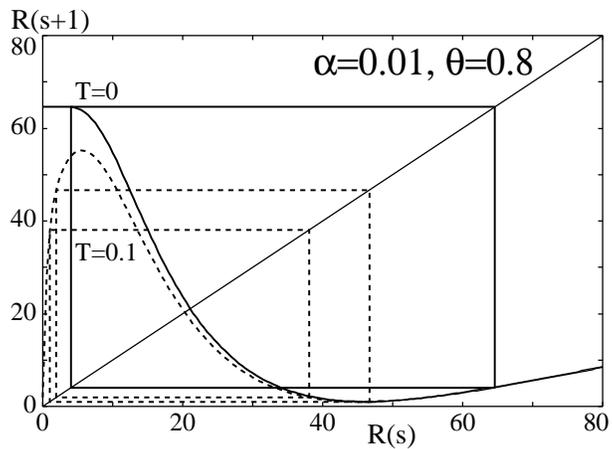}
 \end{center}
 \caption{Return map of $R(t,t)$ for $\alpha=0.01, \theta=0.8$. Solid
 and broken lines denote cases of $T=0$ and $T=0.1$, respectively.}
 \label{fig:mapT}
\end{figure}

Next we discuss the reason for the change in the bifurcation structure
due to the finite temperature effect. We consider the case of an
attractor on $m(t)=0$. Since fixed point $Q$ is on invariant line
$m=0$, the dynamics obeys a one-dimensional map with respect to
$R(t,t)$. 
Figure~\ref{fig:mapT} shows the return map of $R(t,t)$.  For $T=0$
(solid line), the map is unimodal and there is a period-$2$ attractor,
whereas the map is bimodal and there is a period-$4$ attractor for
$T=0.1$ (broken line).  Therefore, the finite temperature effect changes
the bifurcation structure, causing a bistable region to appear. When $T$
is large enough, the more-than-two-period attractors and chaotic
attractors disappear. Thermal noise, therefore, orders the system. The
ordering mechanism may be similar to that of noise-induced order
\cite{MatsumotoTsuda1983,ShibataChawanyaKaneko1999}.

In summary, we considered a sequential associative memory model
consisting of non-monotonic units, which is a stochastic system, and
derived macroscopic state equations using the path-integral method in
the frustrated case. The results obtained by theory agreed with the
results obtained by computer simulation. We constructed two-parameter
bifurcation diagrams for various temperatures and used them to explain
the changes in the structure of the bifurcations caused by the
temperature effect and the coexistence of attractors.

\section*{Acknowledgements}

This work was partially supported by Grant-in-Aid for Scientific
Research on Priority Areas No. 14084212 and Grant-in-Aid for
Scientific Research (C) No. 14580438.

\newpage

%%%%% Figure

%%%%% Figure

%%%%% Figure

%%%%% Figure

%%%%% Figure

%%%%% Figure

%%%%% Figure

%%%%% Figure

%%%%% Figure

%%%%% Figure

%%%%% Figure

\end{document}